\documentstyle[twoside,fleqn,espcrc2]{article}

    \newcommand{\gbar}{\overline{g}}
    \newcommand{\gMSbar}{g_{\overline{\rm MS}}}
    \newcommand{\alphat}{\tilde{\alpha}_0}
    \newcommand{\re}{{\rm e}}
    \newcommand{\rD}{{\rm D}}
    \newcommand{\rO}{{\rm O}}
    \newcommand{\be}{\begin{equation}}
    \newcommand{\ee}{\end{equation}}

\newcommand{\AmS}{{\protect\the\textfont2
  A\kern-.1667em\lower.5ex\hbox{M}\kern-.125emS}}

\title{Two-loop computation of a finite volume running
       coupling on the lattice}

\author{Ulli Wolff\address{Institut f\"ur Physik, Humboldt Universit\"at \\
        Invalidenstr. 110, D-10099 Berlin, Germany}}

\begin{document}

\begin{abstract}
In pure SU(2) gauge theory we compute the two-loop coefficient in the relation
between the lattice bare coupling and the running coupling defined through the
Schr\"odinger functional.  This result is required
to relate the latter to the $\overline{\rm MS}$-coupling in our programme
to compute $\alpha_s$.
In addition it allows us to
implement O(a) improvement of the  Schr\"odinger functional to two-loop order.
The two-loop $\beta$-function is verified in a perturbative
computation on the lattice, and the behavior of expansions in the standard
and in the
Parisi-improved bare couplings
are investigated beyond one loop.
\end{abstract}

\maketitle

\section{CONTEXT OF THE CALCULATION}

In this contribution we report on a perturbative 2-loop
computation of the running coupling constant $\gbar^2(L)$, defined through the
Schr\"odinger functional \cite{schroe}, in terms of the lattice bare
coupling $g_0$.
Such a result is required as part of our project of computing $\alpha_s$
for QCD. The strategy there is to first numerically construct
the continuum limit of
$\gbar$ for some
large values of $L^{-1}$ in physical units.
These are then connected with the more conventional $\gMSbar(q)$
by renormalized perturbation theory. This is expected to be a
well-behaved application of the series,
as the scales $L^{-1}$ and $q$ are of the same order and can be chosen
well above
the nonperturbative scales. With growing statistical accuracy, especially
on parallel computers, the 2-loop coefficient in this conversion is
required to achieve a reasonable balance and  control of the systematic errors
from truncating the series.
The connection between $\gbar$ and $\gMSbar$ is established by expressing
both of them in terms of the lattice bare coupling $g_0$.
In this way,
apart from avoiding problems of a direct connection via
dimensional  regularization in the presence of
a finite background field,
we have the additional benefit of getting the two-loop O($a$) improvement
coefficient to speed up the continuum limit \cite{alpha}.
Here we present the results of a first such
computation of $\gbar$ for SU(2) pure
gauge theory \cite{NW}. For the connection with $\gMSbar$ see Peter Weisz's
contribution to
these proceedings.

\section{FINITE VOLUME RUNNING COUPLING}

The definition of $\gbar$ from the Schr\"odinger functional is discussed
in more detail in \cite{schroe,su2}, and we are brief here.
The effective action $\Gamma$ is given by the path integral
over SU(2) gauge fields $U(x,\mu)$
on a lattice with extension $L$ in all four directions,
\be
  \re^{-\Gamma} = \int \rD[U] \; \re^{-S(U)}.
  \label{pathint}
\ee
While space is periodic,
the fixed boundary values in the time direction,
$U(x,k)|_{x^0=0} = \exp\{\eta \tau_3/ i L\}$ and
$U(x,k)|_{x^0=L} = \exp\{(\pi-\eta) \tau_3/ i L \}$,
induce an $\eta$-dependent abelian background field.
The action $S$ is the standard Wilson action modified
for the surfaces at $x^0 = 0, L$,
\be
  S[U]= {1\over g_0^2} \sum_p w(p)
  {\rm tr} \{1-U(p)\},
\ee
where the sum runs over all oriented plaquettes $p$.
The weight $w(p)$ is unity except for the
the electrical plaquettes touching the fixed
field boundary, where it equals
\be
    c_t(g_0)=1 + c_t^{(1)} g_0^2 + c_t^{(2)} g_0^4 + \cdots.
\ee
Its tuning allows for the suppression of O($a$) artifacts
arising from the presence of the boundaries.
Following Symanzik, we perturbatively determine
the coefficients in
\be
    c_t(g_0)=1 + c_t^{(1)} g_0^2 + c_t^{(2)} g_0^4 + \cdots.
\ee

The response
\be
\bar g^2 (L) =  \left. {\Gamma_0' \over \Gamma' }
                  \right\vert_{\eta = \pi/4}
\ee
defines $\gbar$,
which runs with the finite system size $L$ as the only scale in
the problem. Here $'$ denotes $\eta$-derivatives, and $\Gamma_0$
is the exactly known classical limit ($g_0 \to 0$) of $\Gamma$.

\section{PERTURBATIVE CALCULATION}

A suitable gauge fixing with the introduction of ghosts  and
a systematic expansion of the fluctuations
of the $U$-field around the induced background lead to a
well-defined lattice perturbation expansion \cite{schroe}.
In particular, all modes are
quadratically damped with the present boundary conditions.
In the resulting expansion
\be
 \gbar^2(L) = g^2_0 + m_1(L/a) g^4_0 + m_2(L/a) g^6_0 + \cdots,
 \label{series_gbar}
\ee
the coefficient $m_1$ was given in \cite{schroe} and, at 2 loop order,
$m_2$ is reported here and in more detail in \cite{NW}.
Evaluation of (the $\eta$-dependence of)
the order of 10 vacuum 2-loop diagrams is required for $m_2$.
The computation is involved, because all propagators have to be
computed numerically
with the finite background field in place, and because the
3- and 4-gluon vertices have a large number ($\sim 100$)
of terms. Hence the diagrams are summed numerically
for $L/a \le 32$. To get to this size on workstations in a reasonable time,
all symmetries
were used eventually to reduce the number of terms. On smaller
lattices they were however checked to hold first. As further error checks
independence of a continuous gauge fixing parameter was verified,
and the two authors of \cite{NW} carried out independent calculations
for the smaller lattices before optimizing one of the codes.

The resulting column of values of $m_2$ vs. $L/a$ was analyzed
as described in \cite{perturb}. The asymptotic result is
\begin{eqnarray}
m_1 &=& 2 b_0 \ln(L/a) +0.20235 + \rO(a^2/L^2) \label{coeff_m1}\\
m_2 &=& m_1^2 + \nonumber \\
    && 2 b_1 \ln(L/a) +0.01607 + \rO(a^2/L^2) \label{coeff_m2}.
\label{coeff_gbar}
\end{eqnarray}
Here $b_0, b_1$ are the universal coefficients of the $\beta$-function
\be
  b_0 = {11 \over 24 \pi^2}, \ \ \
  b_1 = {17 \over 96 \pi^4}.
\ee
The value of $b_1$ was confirmed in our analysis of $m_2(L/a)$ to about
1 part in $10^3$, and to our knowledge this is the first such check
on the lattice. The lattice artifacts of O($a$) have been canceled in
$m_1, m_2$. This requirement fixes the improvement coefficients
\be
 c_t^{(1)} = -0.0543(5), \ \ \
 c_t^{(2)} = -0.0115(5).
\ee

\section{EXPANSION IN THE BARE COUPLING AT TWO LOOPS}

As mentioned, the result of the previous section is needed
to connect $\gbar$ and $\gMSbar$ via $g_0$. As a byproduct, we have
the opportunity
to study the quality of the lattice
perturbation expansion in $g_0$ for a physical quantity.
Eqs. (\ref{series_gbar}), (\ref{coeff_m1}), (\ref{coeff_m2}) can be used
to relate $\gbar$ at a scale $a/s, s = \rO(1)$,
 {\em in the continuum} with
the bare lattice coupling $g_0$ associated with the cutoff $a$ in the
regularization in use. On the one hand, no large logarithms are expected
here as scales of the same order are connected. On the other hand,
the expansion involves non-universal lattice quantities and
cannot be systematically improved for fixed physical scale by approaching
the continuum limit.
The precise derivation of the relation, although simple in result,
 is slightly tricky to argue
\cite{alpha}:
One first uses (\ref{series_gbar}) in a regime
close to the continuum limit, where $g_0^2$, $g_0^2 \ln(L/a)$
are small in the expansion, and $a^2/L^2$ is negligible.
The resulting $\gbar^2(L)$ is re-expanded in terms of $\gbar^2(a/s)$
by using continuum perturbation theory. This procedure is
{\em formally equivalent} to dropping powers of $a/L$ in (\ref{coeff_m1}),
(\ref{coeff_m2}) and then putting $a/L = s$ under the logarithms.

In terms of $\alpha(q) = \gbar^2(L)/(4 \pi),\; L=q^{-1}$, and
$\alpha_0 = g_0^2/(4 \pi)$ one gets for $s=1$
\be
\alpha(a^{-1}) = \alpha_0 + 2.543 \, \alpha_0^2 + 9.00 \, \alpha_0^3
  +\rO(\alpha_0^4).
\ee
The rather large coefficients here are consistent with the experience,
that $g_0$ used in this way is not a good expansion parameter.
An alternative way of using the series is to fix $s$ such that
there is no 1-loop term, and this leads to
\be
  \alpha (8.83 \, a^{-1}) = \alpha_0 + 1.287(1) \alpha_0^3 +\rO(\alpha_0^4).
  \label{s883}
\ee
Note that this choice simultaneously produces a small 2-loop coefficient,
which is nontrivial and leaves the chance, that the series may be better
behaved with this rather large scale shift $s=8.83$.

Recently the use of
modified bare couplings has become popular
to try to amend the bare series \cite{LMDallas}.
A simple proposal is \cite{Parisi}
\be
\alphat = \alpha_0/P,
\ee
where $P$ is the average plaquette in infinite volume.
Using the known series for $P$, one finds analogously to (\ref{s883})
\be
  \alpha(1.17 \, a^{-1}) = \alphat + 0.951(1) \alphat^3
  +\rO(\alphat^4).
\label{s117}
\ee
Here the 2-loop coefficient is reasonable also, and the scale factor
is much closer to one.

We now want to compare the expansions (\ref{s883}) and (\ref{s117})
with nonperturbative results for $\gbar$.
We consider $\beta=2.85$, where $\alpha_0$ and $\alphat$
are available, and the scale $a^{-1}_{2.85}$ is about 8 GeV if the SU(2)
potential is matched with nature \cite{alpha}.
The expansions produce $\gbar$
at $1.17\, a^{-1}_{2.85}$ and  $8.83\, a^{-1}_{2.85}$, which, for the
comparison,
we both evolve to $10\, a^{-1}_{2.85}$
with the numerically controlled $\beta$-function of $\gbar$ and
negligible error. At this scale $\gbar^2$ itself is known numerically
\cite{alpha},
and all values are collected in table~\ref{tab}.
\begin{table}[hbt]
\setlength{\tabcolsep}{1.5pc}
\newlength{\digitwidth} \settowidth{\digitwidth}{\rm 0}
\catcode`?=\active \def?{\kern\digitwidth}
\caption{Estimates for $\alpha(10\, a^{-1}_{2.85})$}
\label{tab}
\begin{tabular}{lc}
\hline
$\hspace{1 em} \alpha$ & method \\
\hline
0.1135(8) & nonperturbative \\ [0.5ex]
0.1098    & 1-loop in $\alpha_0$ \\
0.1115    & 2-loop in $\alpha_0$ \\
0.1110    & 1-loop in $\alphat$ \\
0.1128    & 2-loop in $\alphat$ \\
\hline
\end{tabular}
\end{table}
The 2-loop results are only off by about 1  and 2.5 error margins
for $\alphat$ and $\alpha_0$ respectively.

When using the 1-loop term to fix the scale for which to apply
bare perturbation theory, we find both a reasonably  sized 2-loop term
and approximate agreement with the ``exact'' result.
Perhaps a bit surprisingly,
this works for the ordinary bare lattice coupling
in a qualitatively similar fashion as for $\alphat$,
with just the scale factor
being rather large. One has to be cautious, however, that this experience
depends on the renormalized quantity (here $\gbar$) that is
computed and need not be universally true.


\begin{thebibliography}{9}
\bibitem{schroe}M. L\"uscher, R. Narayanan, P. Weisz and U. Wolff,
Nucl. Phys. B384 (1992) 168

\bibitem{su2} M. L\"uscher, R. Sommer, P. Weisz and U. Wolff,
Nucl. Phys. B389 (1993) 247


\bibitem{alpha} G. de Divitiis,
R. Frezzotti,
M. Guagnelli,
M. L\"{u}scher,
R. Petronzio,
R. Sommer,
P. Weisz and
U. Wolff,
{\it Universality and the approach to the continuum limit
in lattice gauge theory},
preprint DESY 94-106, CERN-TH 7447/94, HUB-IEP-94/16,
hep-lat/9411017

\bibitem{NW} R. Narayanan and U. Wolff,
{\it Two loop computation of a running coupling in lattice Yang-Mills theory},
Berlin-Princeton preprint (1994),to appear



\bibitem{perturb} M. L\"uscher and P. Weisz,
Nucl. Phys. B266 (1986) 309.


\bibitem{LMDallas}
A. X. El-Khadra, G. Hockney, A.S. Kronfeld and P. B. Mackenzie,

Phys. Rev. Lett. 69 (1992) 729

G.P. Lepage and  P. B. Mackenzie, Phys. Rev. D 48 (1993) 2250


\bibitem{Parisi} G. Parisi,
{\it in}\/: High-Energy Physics --- 1980,
XX. Int. Conf., Madison (1980), ed. L. Durand and L. G. Pondrom
(American Institute of Physics, New York, 1981)


\end{thebibliography}
\end{document}